\documentclass[twocolumn,prb,aps,showpacs]{revtex4}
\usepackage{epsfig}
\begin{document}
\title{\bf Anomalous Hall Effect in two dimensional paramagnetic systems}
\author{\rm Dimitrie Culcer, Allan MacDonald, Qian Niu}\medskip
\affiliation{\it University of Texas, Austin TX78712}\medskip
\begin{abstract}
We investigate the possibility of observing the anomalous Hall effect (AHE) in two dimensional paramagnetic systems. We apply the semiclassical equations of motion to carriers in the conduction and valence bands of wurtzite and zincblende quantum wells in the exchange field generated by magnetic impurities and we calculate the anomalous Hall conductivity based on the Berry phase corrections to the carrier velocity. We show that under certain circumstances this conductivity approaches one half of the conductance quantum. We consider the effect of an external magnetic field and show that for a small enough field the theory is unaltered.
\end{abstract} 
\maketitle
\section{INTRODUCTION}
When a nonferromagnetic metallic sample is exposed to a perpendicular external 
magnetic field, the Lorentz force acting on the current carriers gives rise to a transverse voltage in the plane of the sample. The transverse component of the resistivity $\rho_{xy}$ depends on the magnetic field through: 
\begin{equation} 
\rho_{xy}=R_0B 
\end{equation} 
where $R_0 = \frac{1}{ne}$ is known as the Hall coefficient. This phenomenon is known as the ordinary Hall effect.

In many ferromagnets, however, the transverse resistivity acquires 
an additional term which is often seen to be proportional to the magnetization of the sample,
and becomes constant once the sample has reached its saturation 
magnetization $M_s$. Empirically one writes: 
\begin{equation}  
\rho_{xy}=R_0B+R_sM 
\end{equation}   
The effect is referred to as the anomalous Hall effect while the constant $R_s$ is called the anomalous Hall coefficient. It can be seen from the second term above that ferromagnets display a spontaneous Hall conductivity in the absence of an external field. The effect was subsequently noted in a large number of bulk alloys, as well as, in recent experiments, in materials which exhibit colossal magnetoresistance$^{1-2}$ and ferromagnetic semiconductors. Recent studies of ferromagnetic semiconductors such as (Ga,Mn)N films have in fact reported ferromagnetic behaviour at room temperature$^{3-4}$.

Although it has been known for close to half a century, the AHE has had a controversial history and it remains a somewhat poorly understood phenomenon. Karplus and Luttinger$^{5}$ pioneered the theory of the AHE, finding that the spin splitting of bands can give rise to a Hall conductivity in the presence of spin orbit coupling. Smit$^{6}$ countered that in a perfectly periodic lattice the AHE could not occur without scattering from impurities, and introduced the skew scattering mechanism to explain it. This mechanism, in which an electron is scattered at an angle to its original direction, gives a contribution proportional to $\rho$, the diagonal resistivity. In a more complete treatment, Luttinger$^{7}$ found a term corresponding to skew scattering but maintained that the scattering free contribution to the AHE still remains. There has been much debate on the possibility of this scattering free contribution in principle and on its relative importance, if it exists, in real materials. 

Later, a new mechanism, called side jump was introduced by Berger$^{8}$ to explain the observed $\rho^2$ dependence, although the scattering free 
contribution gives the same dependence. In side jump, the electron incident into the area of influence of the potential emerges parallel to its original direction but displaced perpendicular to it. This latter term is supposed to dominate in alloys, where $\rho$ is high. It is nevertheless 
not clear how to relate the side jump mechanism to the systematic theory proposed by Luttinger. 

In recent years, the scattering free contribution of Luttinger and Karplus was rederived in a semiclassical analysis of wavepacket motion in Bloch bands by Chang and Niu$^{15}$ and Sundaram and Niu$^{16}$ and was attributed to a Berry phase effect in k-space. A more rigorous derivation$^{13}$ based on the Kubo formula gives the same result. This contribution was also evaluated for the mean-field bands of semiconductor ferromagnets, yielding good agreement with experiments without any parameter fitting. This theory$^{9}$ of the AHE is based on the Stoner description of ferromagnetism, considering the charge carriers to be quasiparticles in spontaneously split Bloch bands. It is to be distinguished from the mechanism of Ye {\it et al} $^{14}$ based on the Berry phase in real space. 
The motivation behind the current effort is to provide a conceptual framework for the theoretical study of the AHE in magnetic quantum wells and heterostructures, which have been realized in recent years. These structures constitute the simplest systems in which the Berry phase can be evaluated analytically from the Hamiltonian including the Rashba spin orbit coupling and provide a suitable ground for testing a theory based on fundamental physics. We shall concentrate our attention on the conduction band and the topmost valence band of an inversion asymmetric semiconductor heterostructure in an exchange field supplied through doping with Mn and calculate the anomalous Hall conductivities for the two bands. Although ferromagnetic behaviour has not been observed in II-VI heterostructures we shall concentrate on II-VI semiconductors, as they can be doped with Mn more heavily than III-V. 

The paper is organized as follows. In section II, within the framework of the effective mass approximation applied to a doubly degenerate band, we calculate the Berry phase of the wavefunction which yields the off-diagonal conductivity $\sigma_{xy}$. We consider an ideal situation with T=0, where we follow the method used by Chang and Niu$^{15}$ and Sundaram and Niu$^{16}$ for the semiclassical treatment of carrier motion in 2D. Under certain circumstances one can make the approximation that the anomalous Hall conductivity is quantized, taking the values: 
\begin{equation} 
|\sigma_{xy}|= \frac{e^2}{2h}
\end{equation}
In sections III and IV we apply the theory to wurtzite and zincblende structures respectively. We consider finite temperature corrections and discuss the conditions under which equation (3) holds. Moreover, we investigate the variation of the conductivity with temperature, exchange coupling and spin orbit constant. In the last section we examine the effect of placing the system in an external magnetic field and determine the optimal parameters needed for the observation of the AHE in heterostructures in the laboratory.

\section{Symmetry considerations}
In a perfect crystal, according to Bloch's theorem, the wavefunction for a band with index n is decomposed into two parts:
\begin{equation}
|\Psi_n ({\bf k}, {\bf r})\rangle = e^{i{\bf k}\cdot{\bf r}}|u_n ({\bf k}, {\bf r})\rangle
\end{equation}
where $|u_n ({\bf k}, {\bf r}) \rangle$ is a function with the periodicity of the lattice. The semiclassical motion of a charge carrier through the crystal is described by constructing a wavepacket out of Bloch wavefunctions. The dynamics of such a wavepacket are given by the following equations of motion$^{16}$:
\begin{eqnarray}
{\bf \dot r} = \frac{1}{\hbar}\frac{\partial \varepsilon_n}{\partial {\bf k}} - {\bf \dot k}\times {\bf \Omega}_n\\
{\bf \dot k} = -\frac{e}{\hbar}({\bf E} + {\bf \dot r} \times {\bf B})
\end{eqnarray}
which determine the position vector and wavevector of the center of the wavepacket in the presence of external electromagnetic fields, with ${\bf \Omega}$ the Berry curvature. The Berry curvature of a band is defined by the following expression:
\begin{eqnarray}
{\bf\Omega}_n = -Im\langle\frac{\partial u_n}{\partial {\bf k}}| \times |\frac{\partial u_n}{\partial {\bf k}}\rangle
\end{eqnarray}
The term containing the Berry curvature is usually neglected, due to the fact that it frequently vanishes by symmetry, as in crystals which are invariant with respect to both time reversal and spatial inversion (e.g. nonmagnetic Bravais crystals$^{16}$).

In the AHE, the additional contribution to the current is perpendicular to the direction of the electric field and independent of the magnetic field. We now show it to be related to the Berry curvature, ${\bf \Omega}_n$ which appears in the equations of motion as an additional term in the velocity. This is the same as the velocity correction derived previously by Luttinger. From the two equations it is apparent that this correction term is perpendicular to ${\bf \dot k}$ and therefore perpendicular to the direction of the Lorentz force. In the absence of an external magnetic field {\bf B}, this term is seen to be perpendicular to the electric field {\bf E}, giving a transverse component of the velocity. This velocity adds a transverse term in the current producing a contribution to the off diagonal conductivity. Therefore, as long as ${\bf \Omega}_n$ is nonzero it is possible to have an off diagonal conductivity term which is independent of {\bf B}. 

The Berry curvature is related to the Berry phase$^{12}$ (denoted by $\gamma_n$), which is the phase acquired by the wavefunction upon being transported around a loop in k-space. According to Stokes' theorem:
\begin{equation}
\int_A d{\bf S}\cdot{\bf \Omega}_n = \int_{\partial A} d{\bf k}\cdot\langle u_n| \frac{\partial}{\partial {\bf k}}| u_n \rangle = \gamma_n
\end{equation}
In the above the loop around which the wavefunction is transported is denoted by $\partial$A and the area enclosed by the loop by A. The Berry curvature can therefore be regarded as the Berry phase per unit area of k-space.

In the following we give an analysis of the main symmetry aspects of the problem. From the requirement that the semiclassical equations be invariant under time reversal it is apparent that ${\bf \Omega}_n$ must be odd under this transformation, namely {${\bf\Omega}_n$(-{\bf k})}=-{${\bf\Omega}_n$({\bf k})}. A geometric argument can also be made by noting that the Berry phase is a path dependent quantity. Under time reversal, both the path along which the wavefunction is transported and the orientation of the wavevector {\bf k} are reversed. A clockwise path spanning a set of wavevectors $\{{\bf k}\}$ becomes an anticlockwise path spanning the set of wavevectors $\{-{\bf k}\}$. This implies that the Berry phase changes sign uner time reversal and the Berry curvature satisfies the above constraint.

One can also obtain this result by carrying out the explicit transformation of Eq.(7) under time reversal. If $|u_n \rangle$ is written in terms of the real and imaginary parts of its components:
\begin{equation}
|u_n \rangle = \left (\matrix{Re |v_n\rangle + iIm |v_n\rangle \cr Re |w_n\rangle + iIm |w_n\rangle \cr}\right)
\end{equation}
then application of the time reversal operator will result in:
 \begin{equation}
{\cal T}|u_n \rangle = \left (\matrix{-i Re |w_n\rangle - Im |w_n\rangle \cr i Re |v_n\rangle + Im |v_n\rangle \cr}\right)
\end{equation}
producing a change of sign in the Berry curvature. 

If time reversal symmetry is present, Kramers degeneracy must also be present, imposing $\varepsilon_n$({\bf k}) = $\varepsilon_n$(-{\bf k}). Therefore, if the state at wavevector {\bf k} is occupied so is the state at wavevector -{\bf k}. This, together with the condition {${\bf\Omega}_n$(-{\bf k})}=-{${\bf\Omega}_n$({\bf k})} implies that the integral of ${\bf\Omega}_n$ over all filled states vanishes. Therefore, in general it is always necessary for the system to lack time reversal symmetry in order for the AHE to occur. 

To obtain a nonzero anomalous Hall conductivity, the spin orbit interaction must also be present in order to couple the spin up and spin down bands. This coupling transfers the time reversal violation from the spin degree of freedom to the orbital motion, which is responsible for the Berry curvature. An example is provided by ferromagnetic (Ga,Mn)As$^{10}$ crystals, in which, without spin orbit, the valence band wavefunctions at k=0 are eigenstates of ${\bf \hat L}$, the orbital angular momentum operator, with eigenvalue l=1 and thus sixfold degenerate. When spin orbit is included the k=0 band wavefunctions are eigenstates of the total angular momentum operator ${\bf \hat J}$, splitting into a fourfold degenerate j=3/2 level (containing the heavy holes and the light holes) and a twofold degenerate j=1/2 level (the split off band). Away from k=0, there is a correction proportional to $({\bf \hat J}\cdot{\bf k})^2$, which partially lifts the degeneracy of the bands. This term provides a k-dependent quantization direction for the angular momentum, so that as the wavevector is displaced the angular momentum is rotated and it is possible to obtain a nonzero Berry curvature. 

In 2D, the quantum confinement lifts the degeneracy of the heavy hole and light hole bands, so that at k=0 it is possible to separate the Hamiltonian into independent 2$\times$2 blocks. For finite k, each block remains degenerate in the presence of both time reversal and spatial inversion symmetries, based on Kramers' theorem. With time reversal symmetry $|{\bf k}, \uparrow \rangle$ is equivalent to $|-{\bf k}, \downarrow \rangle$ while with space inversion symmetry $|{\bf k}, \uparrow \rangle$ is equivalent to $|-{\bf k}, \uparrow \rangle$. Therefore, with both symmetries $|{\bf k}, \uparrow \rangle$ is equivalent to $|{\bf k}, \downarrow \rangle$ . In the absence of space inversion symmetry it is possible to break the degeneracy at each finite k. The space inversion asymmetry gives rise to the Rashba spin orbit interaction$^{21}$:
\begin{equation} 
V_{so}=\alpha_{mat} f(k) (\bf \sigma \times {\bf k})\cdot {\bf \hat z}  
\end{equation} 
where $\alpha_{mat}$ is a constant, $\bf{\sigma}$ is the vector of Pauli spin matrices, k is the two dimensional wavevector in the $xy$ plane and f(k) depends only on the magnitude of the wavevector. The asymmetry can originate from either the crystal structure (bulk inversion asymmetry) or the confinement potential (structure inversion asymmetry). The Rashba interaction has been found to be the main mechanism responsible for the zero field spin splitting in 2DEGs$^{17-20}$. 

It is apparent from the above that the spin orbit coupling provides a k-dependent quantization direction for the charge carriers' spins. The spins prefer to lie in the $xy$ plane and be perpendicular to the wavevector. As a result, when the wavevector sweeps a circle around the origin, the spins are rotated by a solid angle of 2$\pi$, and acquire a Berry phase of $\pi$. Since this phase is independent of the area enclosed, it follows that the Berry curvature is singular at the origin and is null everywhere else.
When an exchange field is applied, the spins are tilted out of the $xy$ plane. The amount of tilting depends on the competition between the Rashba term and the exchange field. From the k-dependence of the Rashba term it can be seen that the solid angle swept by the spins is different from 2$\pi$ and depends on the size of k, tending to zero as the radius of the circle tends to zero. This implies that the Berry curvature is now spread out and is finite at the origin. As will be shown in more detail in the following sections, such a Berry curvature will lead to a finite contribution to the AHE.

\section{General treatment of Berry curvature and Hall conductivity} 

The 2D anomalous Hall conductivity is calculated  at T=0 and shown to be quantized. We consider a $2\times 2$ Hamiltonian describing the spin split conduction/valence band of a semimagnetic semiconductor in the presence of an exchange field and spin orbit coupling. Effects of band mixing are neglected, which is a suitable approximation for the bandstructures of the materials we shall consider-the top two valence bands and the conduction band in wurtzite structures and the conduction band of zincblende materials. 

The exchange field due to the magnetic impurities is taken to be uniform and directed along the $z$ axis, normal to the heterostructure. Based on a mean field model, we consider the interaction to be described by a vector ${\bf h}_0$, which for simplicity has units of energy. The magnitude of the interaction is tuned by controlling the concentration of Mn but its effect will be masked by thermal fluctuations once h$_0 \le$ k$_B$ T. 

In a narrow quantum well in which the subbands are widely separated the ${\bf k\cdot p}$ Hamiltonian, with $m^*$ the band electron effective mass, $\gamma = \frac{\hbar^2}{2m^*}$ and k$_\pm$=k$_x\pm$ik$_y$ is: 
\begin{equation}
H=\gamma k^2 I_{2\times2} +
\left(\matrix{h_0&i\alpha_{mat}f(k)k_-\cr                                 
   -i\alpha_{mat}f(k)k_+&-h_0\cr}\right)  \end{equation}
It is readily seen to have the eigenvalues:
\begin{equation}
E_{\pm}= \gamma k^2 \pm \sqrt{h_0^2 + \alpha_{mat}^2 k^2 f(k)^2}
\end{equation}
yielding two subbands, separated by 2h$_0$. 

Assuming T=0 for the time being we take the bottom subband to be 
occupied and the top one to be empty, while the Fermi level corresponds
to k$_F$=(4$\pi$n)$^{1/2}$. 

The form of the Berry curvature for a general f(k) is:
\begin{equation}
\Omega_z^{\uparrow/\downarrow} = \mp\frac{1}{2} \frac{\alpha_{mat}^2h_0 f(k) \frac{d}{dk}[kf(k)]}{[h_0^2 + \alpha_{mat}^2k^2f(k)^2]^{3/2}} 
\end{equation}

The geometrical phase factor is the integral of the curvature over all wavevectors$^{13}$. As the upper band is empty the integral over it is zero and one only needs to consider the curvature of the lower band, $\Omega^\downarrow$:
\begin{equation}
\Gamma^\downarrow = \int \!\!\! \int_{k<k_F} {\Omega^\downarrow}_z d^2{\bf k}   = \pi[1-\frac{h_0}{[h_0^2+\alpha_{mat}^2k_F^2f(k_F)^2]^{1/2}}]
\end{equation}
To maximize the conductivity, the interval between k = 0 and k = k$_F$ should cover the region over which the Berry curvature is significant, so that k$_F$
must be equal to several times k$_c$, the wavevector at which the curvature falls to half its maximum value. As k$_F$ is fixed by the number density the way to accomplish this is to have h$_0<\!\!<\alpha_{mat}$ k$_F$. When this relation holds the phase $\Gamma^\downarrow$ is very nearly $\pi$.

At zero temperature the conductivity $\sigma$ for a full band is equal to the integral over the Brillouin zone of the component of the Berry curvature parallel to $z$ and is thus proportional to the Berry phase. The upper limit of the integral can be taken to infinity:  
\begin{equation}
\sigma_{xy}^\downarrow = - \frac{e^2}{h} \int\!\!\!\int_{k_F \rightarrow \infty}
{\Omega}_z^\downarrow \frac{d^2{\bf k}}{2\pi}
\end{equation}
which results in:
\begin{eqnarray}
\sigma_{xy}=  -\frac{e^2}{2h} \int_0^{\infty}dk \frac{\alpha_{mat}^2h_0kf(k) \frac{d}{dk}[kf(k)]}{[h_0^2 + \alpha_{mat}^2k^2f(k)^2]^{3/2}} = -\frac{e^2}{2h}
\end{eqnarray}
From the above we see that the conductivity is approximately quantized, regardless of the form of f(k). It is worth noting that $\sigma$ does not depend on the size of the spin orbit splitting constant $\alpha_{mat}$ nor on the magnitude of the external magnetic field and that exact quantization occurs when the Berry phase is $\pi$, i.e. the spin lies in the $xy$ plane.

\section{Wurtzite structures at finite temperatures}
The conduction band and the bottom valence band of wurtzite transform according to the $\Gamma_7$ representation of the rotation group at k=0, while the top valence band transforms according to $\Gamma_9$. The latter however is known empirically not to exhibit a linear spin splitting. 
 
The coefficient $\alpha_{mat}$ introduced above is replaced by $\alpha_{w}$. Then the interaction for the $\Gamma_7$ band is given by:   
\begin{equation} 
V_{so}=\alpha_w (\bf \sigma \times {\bf k})\cdot {\bf \hat z} 
\end{equation} 
The energy bands, corresponding to the dispersion relation: 
\begin{equation}
E_{\pm}= \gamma k^2 \pm \sqrt{h_0^2 + \alpha_{w}^2 k^2}
\end{equation}
are plotted in Fig. 1 as a function of k.
\vskip 12pt
\begin{figure}[h]
\begin{center}
\epsfxsize=2.5in
\epsffile{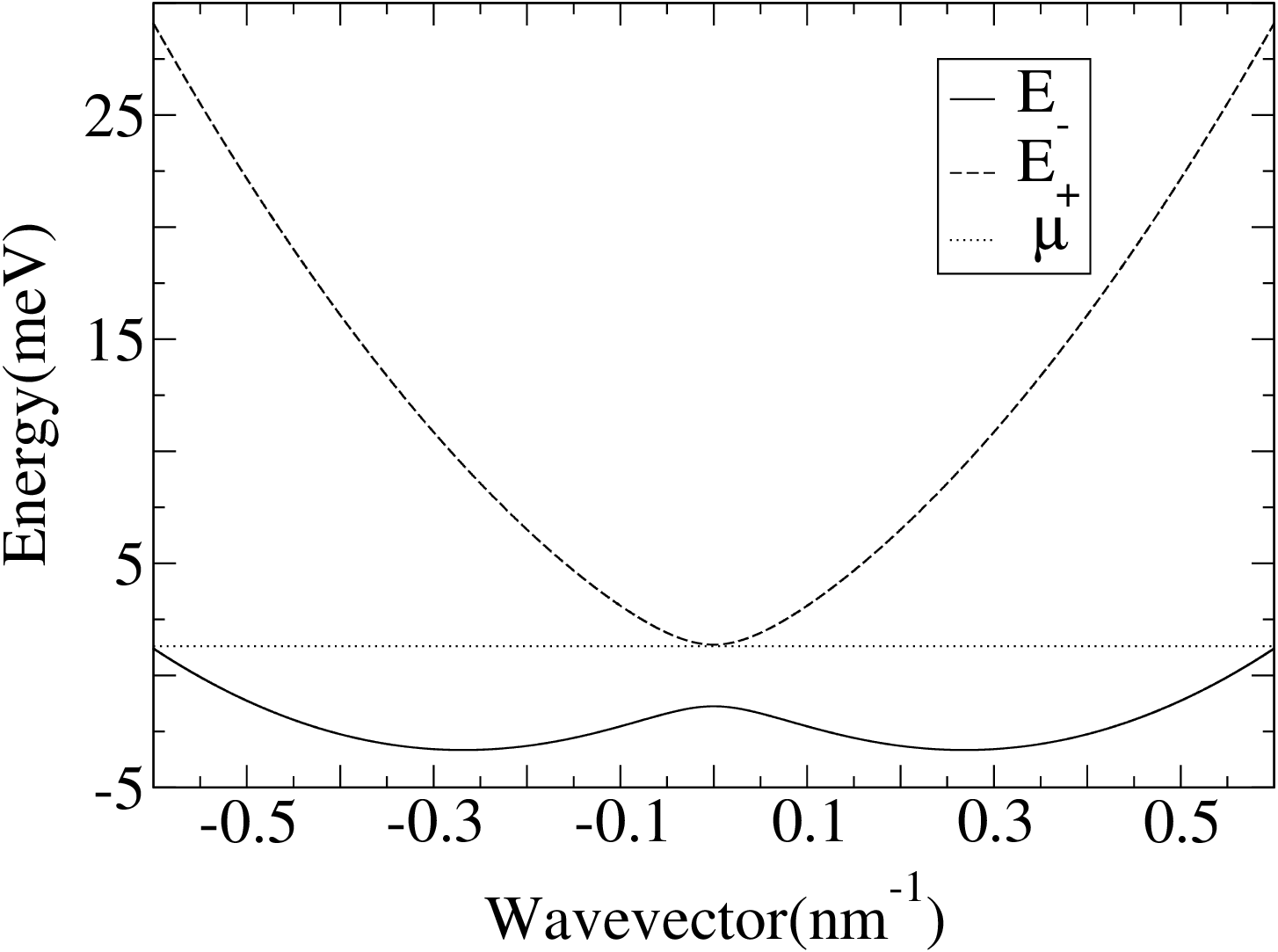}
\vspace{0.2cm} 
\caption{Band dispersion relation for 2D holes in the $\Gamma_7$ valence band of wurtzite structures. The parameters are n=2.9$\times$10$^{12}$cm$^{-2}$, $\alpha_{w}$=23meVnm, h$_0$=1.38meV and m$^*$=0.9m$_0$.}
\label{fig: energiesvb}
\end{center}
\end{figure}
The Berry curvature is pointing along the $z$ axis:
\begin{equation}
\Omega_z^{\uparrow/\downarrow} = \mp \frac{1}{2}\frac{\alpha_w^2h_0}{({\alpha_w^2k^2 + h_0^2})^{3/2}} 
\end{equation}
\vskip 12pt
\begin{figure}[h]
\begin{center}
\epsfxsize=2.5in
\epsffile{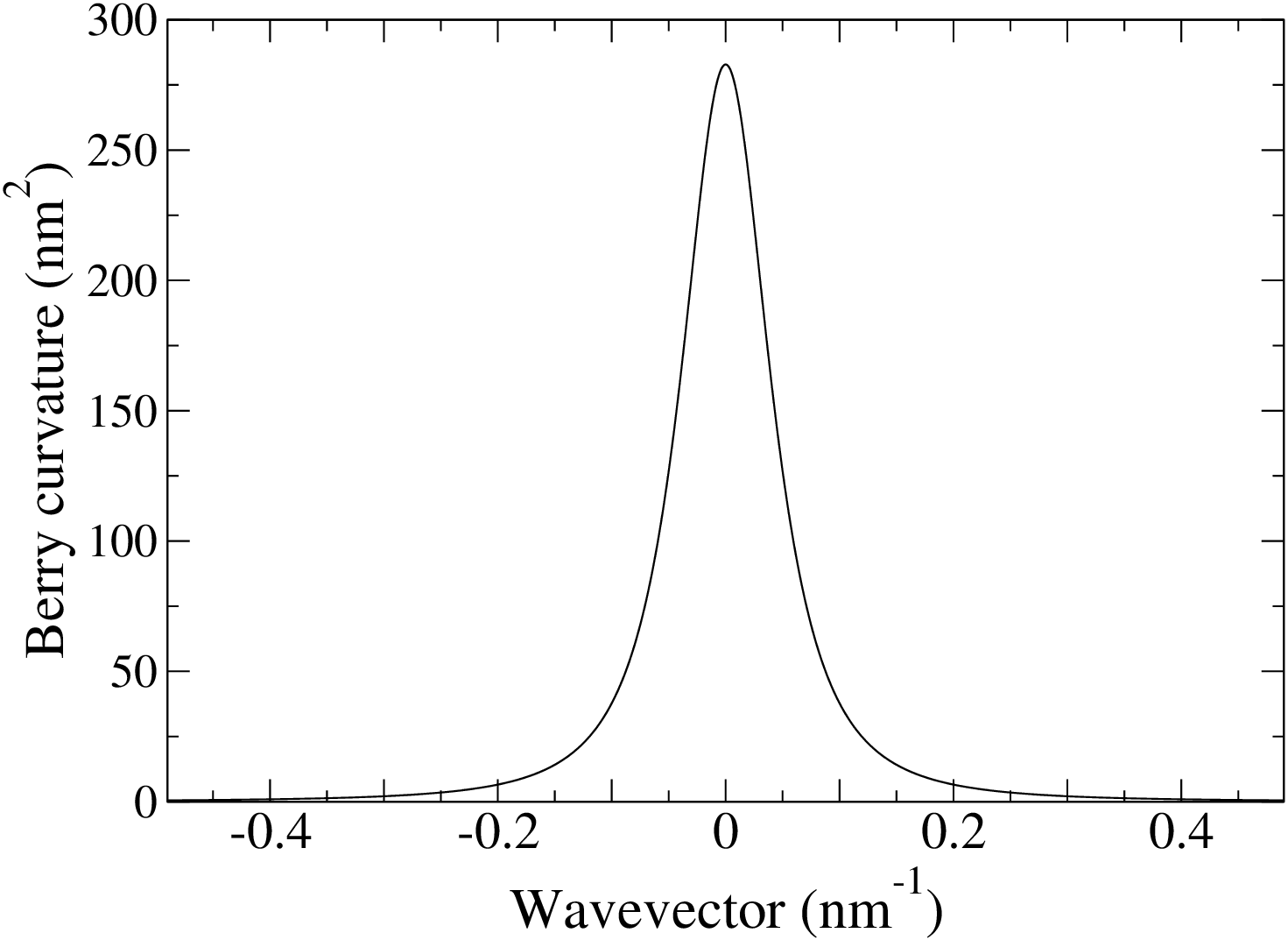}
\vspace{0.2cm} 
\caption{Absolute value of Berry curvature $\Omega_z$ as a function of wavevector for the $\Gamma_7$ valence band of wurtzite structures.}
\label{fig: omegavb}
\end{center}
\end{figure}
The absolute value of the Berry curvature, $\Omega_z$ is plotted in Fig. 2. It falls to half its maximum value when the wavevector is equal to:
\begin{equation}
k_c= \pm \frac{0.77 h_0}{\alpha_w} 
\end{equation}
and its effect becomes negligible once the magnitude of k exceeds several times that of k$_c$. 

At finite temperatures one must take into account the fact that the Fermi-Dirac distribution deviates from the step function at T=0, which is done by incorporating the distribution function into the expression for $\sigma_{xy}$. It is also important to maintain a carrier number density in the range in which AHE is not overshadowed by disorder effects. High densities cause interface effects to become important whereas low densities will cause pockets of electrons to be isolated in localized states. In addition to the above, one must consider the contribution from both the lower and the upper band as there exists a finite fraction of carriers excited into the E$_+$ band. The two conductivities are:
\begin{equation}
\sigma_{\uparrow/\downarrow}= \pm \frac{e^2}{2h} \int_0^{\infty}dk \frac{k \alpha_w^2h_0}
{(h_0^2 + \alpha_w^2k^2)^{3/2}}\frac{1}{e^{(E_{\pm}(k)-\mu)/{k_B T}}+1}
\end{equation}
with E(k) given by (7). The total conductivity $\sigma_{xy}$ is the sum of 
the two:
\begin{equation}
\sigma_{xy}= \sigma^\uparrow_{xy}+\sigma^\downarrow_{xy}
\end{equation}
We consider the conduction band and concentrate on CdSe, where $\alpha_{w}$ has been measured to be 10meVnm and the effective mass m$^*$ is 0.13m$_0$. $\mu$ is determined by the number density and exchange field, which are fixed at $1\times10^{11}$ cm$^{-2}$ and 0.8meV. Our numerical calculations show that under these conditions the maximum conductivity is:
\begin{equation}
|\sigma_{xy}| = 0.125 \frac{e^2}{h}
\end{equation}
It is not quantized, but the effect is still observable.
\vskip 12pt
\begin{figure}[h]
\begin{center}
\epsfxsize=2.5in
\epsffile{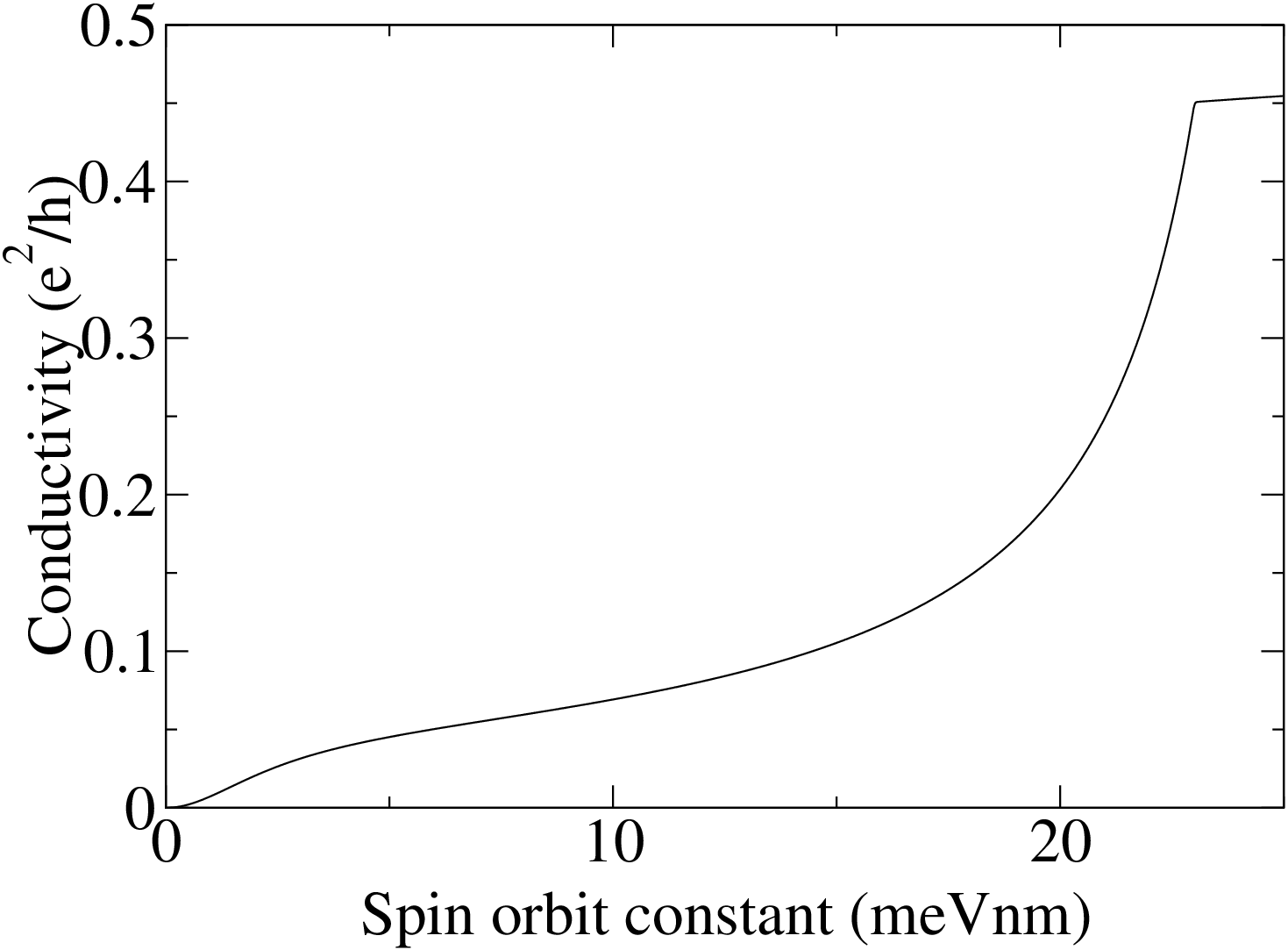}
\vspace{0.2cm} 
\caption{Variation of the conductivity with $\alpha_{w}$ in the case of the $\Gamma_7$ valence band of wurtzite}
\label{fig: asigmavb}
\end{center}
\end{figure}

In the case of the valence band, theory gives an estimate for $\alpha_w$
of 23 meVnm while experiment sets an upper limit of 90 meVnm, and we employ the theoretical value as a worst case scenario. The effective mass is 0.9m$_0$, the number density is set to $2.9\times10^{12}$ cm$^{-2}$, h$_0$ is fixed at 1.38meV and the temperature at 0.1K. Repeating the calculation yields:
\begin{equation}
|\sigma_{xy}| = 0.45 \frac{e^2}{h}
\end{equation}
showing that the conductivity approaches the quantized value.

We now investigate the dependence of the integral in equation (16) upon the spin orbit coupling constant (Fig. 3), maintaining the other parameters at their values for the valence band. We find that at T=0.1K it increases with increasing $\alpha_w$, saturating to $0.45\frac{e^2}{h}$. The shape of the graph can be explained by noting that the effect of increasing $\alpha_w$ is to bring down the chemical potential and flatten the lower band in such a way that its Fermi wavevector is unchanged. The point where the conductivity reaches its maximum corresponds to the point where the chemical potential crosses from the top band into the bottom one so that at very low temperatures only the latter is occupied. Since there are only carriers in the lower band, as $\alpha_w$ increases they acquire approximately the same Berry phase until the chemical potential touches the band maximum at k = 0, beyond which our theory does not apply. The shape of the curve as far as the plateau follows from the fact that as the chemical potential is lowered fewer states are available in the upper band. The plateau itself is understood by noting that increasing $\alpha_w$ makes the curvature narrower but after a point almost all the area over which $\Omega$ is appreciable has been covered, so further increasing $\alpha_w$ will not make a considerable difference. 
\vskip 24pt
\begin{figure}[h]
\begin{center}
\epsfxsize=2.5in
\epsffile{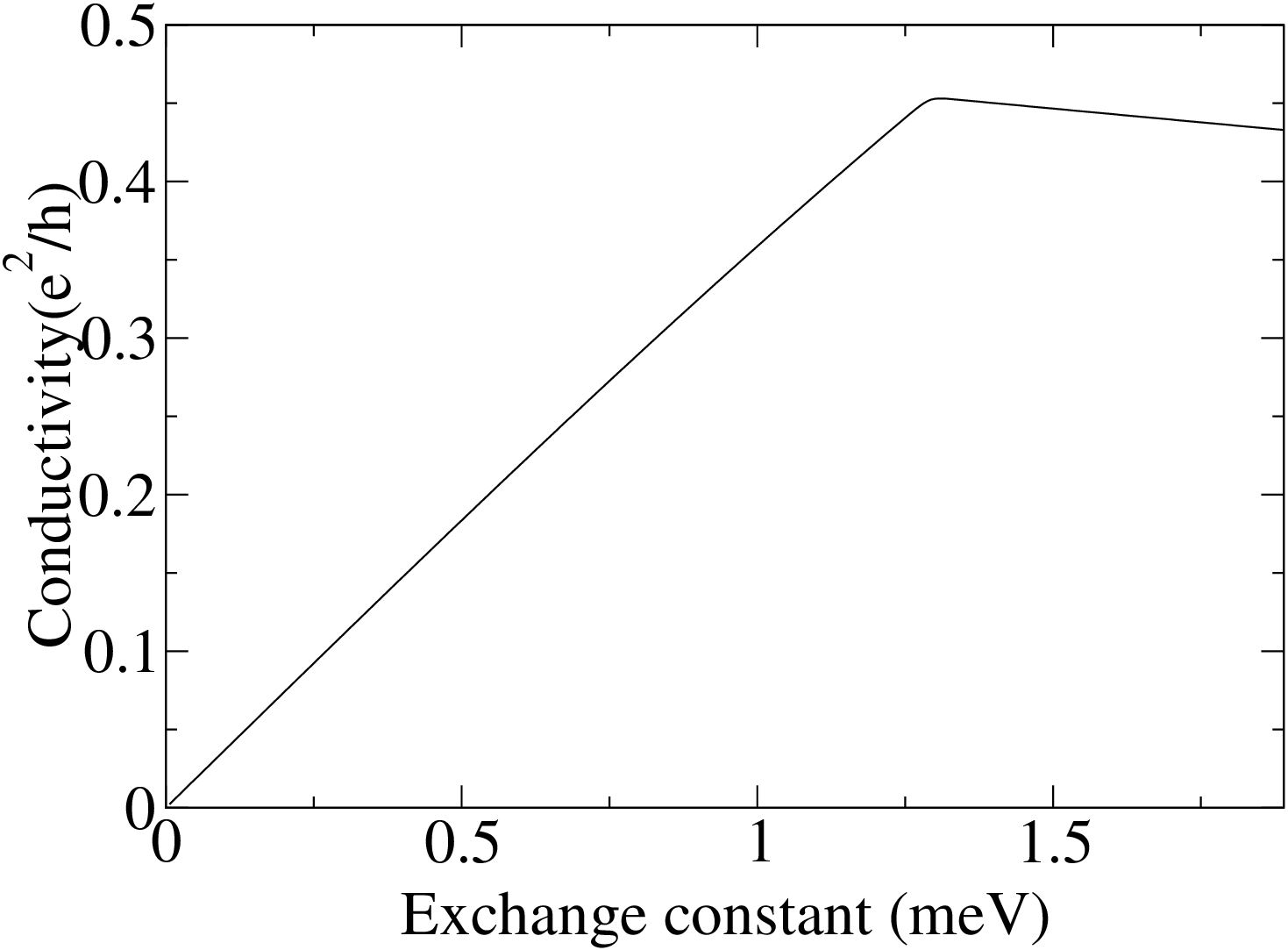}
\vspace{0.2cm} 
\caption{Variation of the conductivity with $h_0$ for the $\Gamma_7$ valence band of wurtzite}
\label{fig: h0sigmavb}
\end{center}
\end{figure}

The dependence upon the exchange coupling is studied next (Fig. 4). It can be seen that $\sigma_{xy}$ reaches a maximum when h$_0$ is approximately 1.38 meV, after which it drops. At first, when there is no magnetic interaction, the spin lies in the $xy$ plane. As h$_0$ increases the spin is tilted out of the plane by larger amounts, increasing the phase acquired by the wavefunction, until it reaches a maximum. As h$_0 \rightarrow \infty$ the spin becomes parallel to h$_0$ and the phase gradually falls to zero. Increasing $h_0$ makes $\Omega$ wider so less of the curvature is covered in the range k = 0 to k$_F$. The sudden fall in the conductivity beyond the maximum is therefore a combined effect - the magnitude of the curvature is smaller and less of the curvature is covered in the integral.

These two plots illustrate the fundamental physics of the system, namely the interplay between the Rashba and Zeeman effects giving rise to the 
anomalous Hall conductivity through the Berry phase acquired by the wavefunction. The dynamics can be viewed as a competition between the Zeeman term, which by itself would align the spin with the $z$ axis and the Rashba term, which draws it towards the $xy$ plane. Without the spin orbit interaction, the Berry phase is zero yielding zero conductivity whereas without the exchange field the energy gap vanishes and the bands overlap. What is more, as h$_0$ tends to infinity the spins align themselves along $z$ in such a way that the wavefunction does not acquire a Berry phase. At this stage the wavevector precesses on the Fermi surface at an infinite rate, which is equivalent to no precession at all. Lastly, as $\alpha_w$ tends to zero the spins once more align with the $z$ axis. 

Finally, we have observed the temperature dependence of the integral in
equation (10), with $\alpha_w$ chosen as before. As Fig. 5 shows, the conductivity declines over the range T=10mK to T=1K, which is attributed to the fact that raising the temperature causes more carriers to be excited across the gap, increasing the size of the negative contribution. 

These two situations are similar to the limit h$_0 \rightarrow 0$:
the bandgap here does not disappear, but it is bridged by facilitating
the movement of carriers across it.
\vskip 12pt
\begin{figure}[h]
\begin{center}
\epsfxsize=2.5in
\epsffile{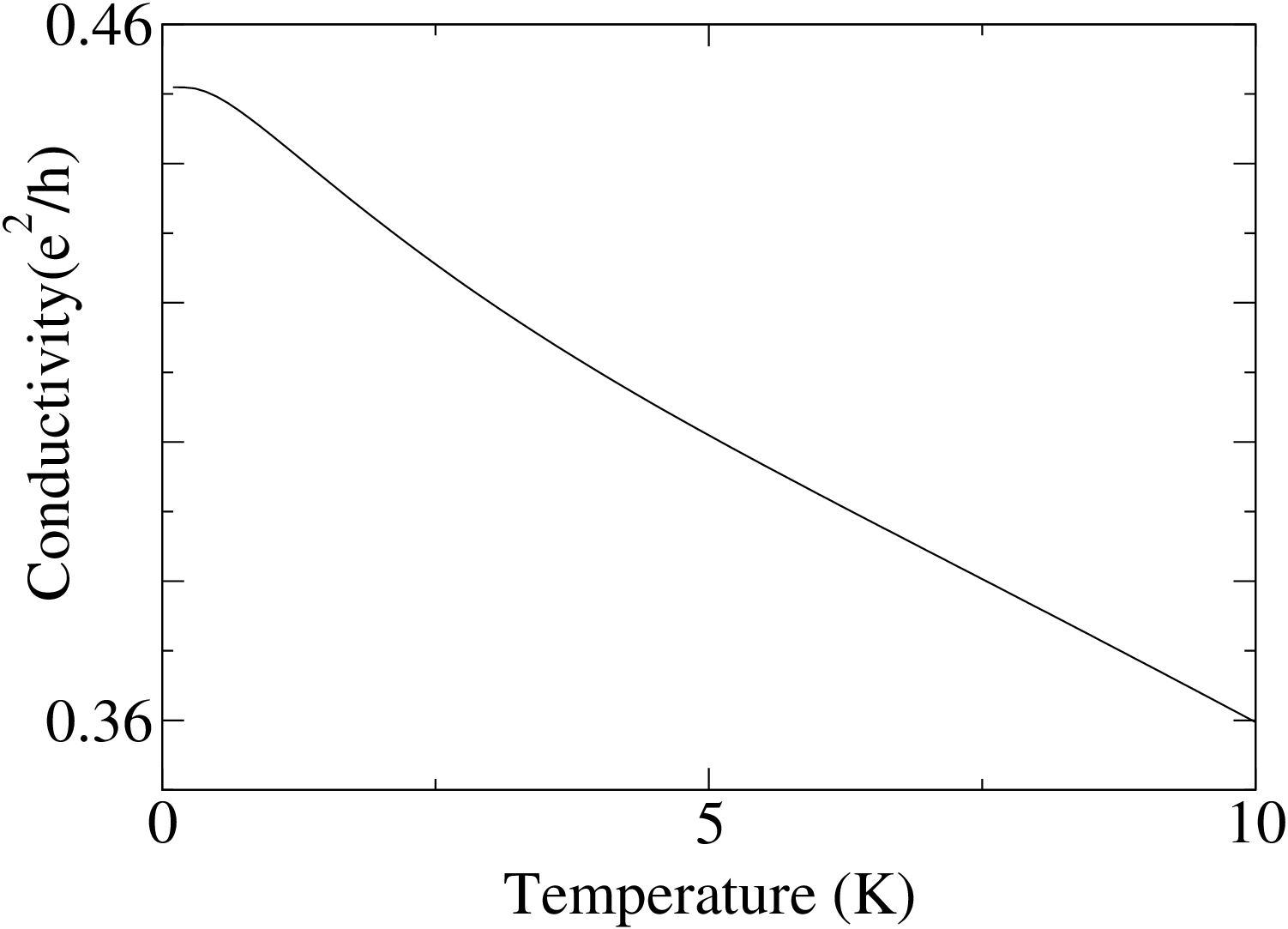}
\vspace{0.2cm}
\caption{Variation of the conductivity with temperature for the $\Gamma_7$ valence band of wurtzite}
\label{fig: tsigma}
\end{center}
\end{figure}  

\section{Zincblende structures at finite temperatures}
Having investigated the underlying physics of the problem for a wurtzite QW, we turn our attention to a case which promises immediate experimental realization. We shall restrict our discussion of zincblende materials to the conduction band of Hg$_{1-x}$Mn$_x$Te, in which the linear term in k is not allowed by symmetry. Instead, the first term in the expansion is cubic in k and the spin orbit term takes the form:
\begin{equation} 
V_{so}=\alpha_{zb}k^2 (\bf \sigma \times {\bf k})\cdot {\bf \hat z} 
\end{equation} 
where $\alpha_{zb}$ replaces $\alpha_{mat}$. This expression is valid near k=0, but is not accurate as k approaches k$_F$. In order to improve the accuracy we have chosen the polynomial coefficients b$_1$ and b$_2$ so as to match the dispersion relation with that shown in Fig. 7 of ref. 37, namely:
\begin{equation} 
V_{so}=\frac{\alpha_{zb}k^2 (\bf \sigma \times {\bf k})\cdot {\bf \hat z}}{1 + b_1k^2 + b_2k^4} 
\end{equation} 
yielding the energy bands (Fig. 6):
\begin{equation}
E_{\pm}= \gamma k^2 \pm \sqrt{h_0^2 + \frac{\alpha_{zb}^2 k^6}{(1 + b_1k^2 + b_2k^4)^2}}
\end{equation}
and the absolute value of the z-component of the Berry curvature (Fig. 7):
\begin{equation}
\Omega_z^{\uparrow/\downarrow} = \mp \frac{\alpha_{zb}^2h_0k^2}{2(h_0^2 + {\alpha_{zb}^2k^6})^{3/2}} \frac{(3k^2 + b_1 k^4 - b_2 k^6)}{(1 + b_1k^2 + b_2k^4)^3}
\end{equation}
We consider the optimum achievable conditions for the observation of the anomalous Hall conductivity. The doping density is n = 2.8$\times$10$^{11}$cm$^{-2}$, the spin orbit constant from ref. 37 is approximately $\alpha_{zb}$ = 10000meV nm$^3$, we set the exchange field to be equal to 3.38meV and m$^*$=0.034m$_0$.
Under these conditions the conductivity is:
\begin{equation}
|\sigma_{xy}|=0.34\frac{e^2}{h}
\end{equation}
In Fig. 7-9 we have plotted the conductivity as a function of the spin orbit constant, exchange field and temperature. The graphs will be seen to have very similar features to the corresponding ones for wurtzite. These common features have identical explanations in terms of the modification of the shape of the bands and the movement of the chemical potential relative to them, as discussed above. 

It will be noticed that the zincblende graphs are smoother and the plateau in the spin orbit constant graph is missing. The qualitative differences in the behaviour of the conductivity come about due to the difference in the shape of the bandstructure and Berry curvature in the two structures. In wurtzite $\Omega$ peaks at the origin and is appreciable within a disc centered at k = 0. In zincblende on the other hand the curvature is zero at the origin and is concentrated within a ring on either side of the values of {\bf k} at which it peaks. If the magnitude of the wavevector at which $\Omega$ has its maximum is denoted by k$_\Omega$, it emerges that in order to maximize the anomalous conductivity the parameters must be adjusted such that k$_F$ is large enough to contain the ring on the outer side of k$_\Omega$ but small enough for the number of states available in the upper band not to cause the contribution from it to cancel out the curvature from the lower band. In the spin orbit constant graph, after reaching a maximum, the conductivity quickly declines, since increasing $\alpha_{zb}$ has the effect of lowering the chemical potential, so that fewer states in the bottom band are integrated over. Due to the shape of the curvature, lowering the chemical potential causes those wavevectors at which the Berry curvature is significant to be omitted, resulting in a sharp decrease in the conductivity. Moreover, the fact that in zincblende the lower band does not have a maximum at k=0 means that our theory can be applied regardless of where in the band the chemical potential lies. 
\vskip 18pt
\begin{figure}[h]
\begin{center}
\epsfxsize=2.5in
\epsffile{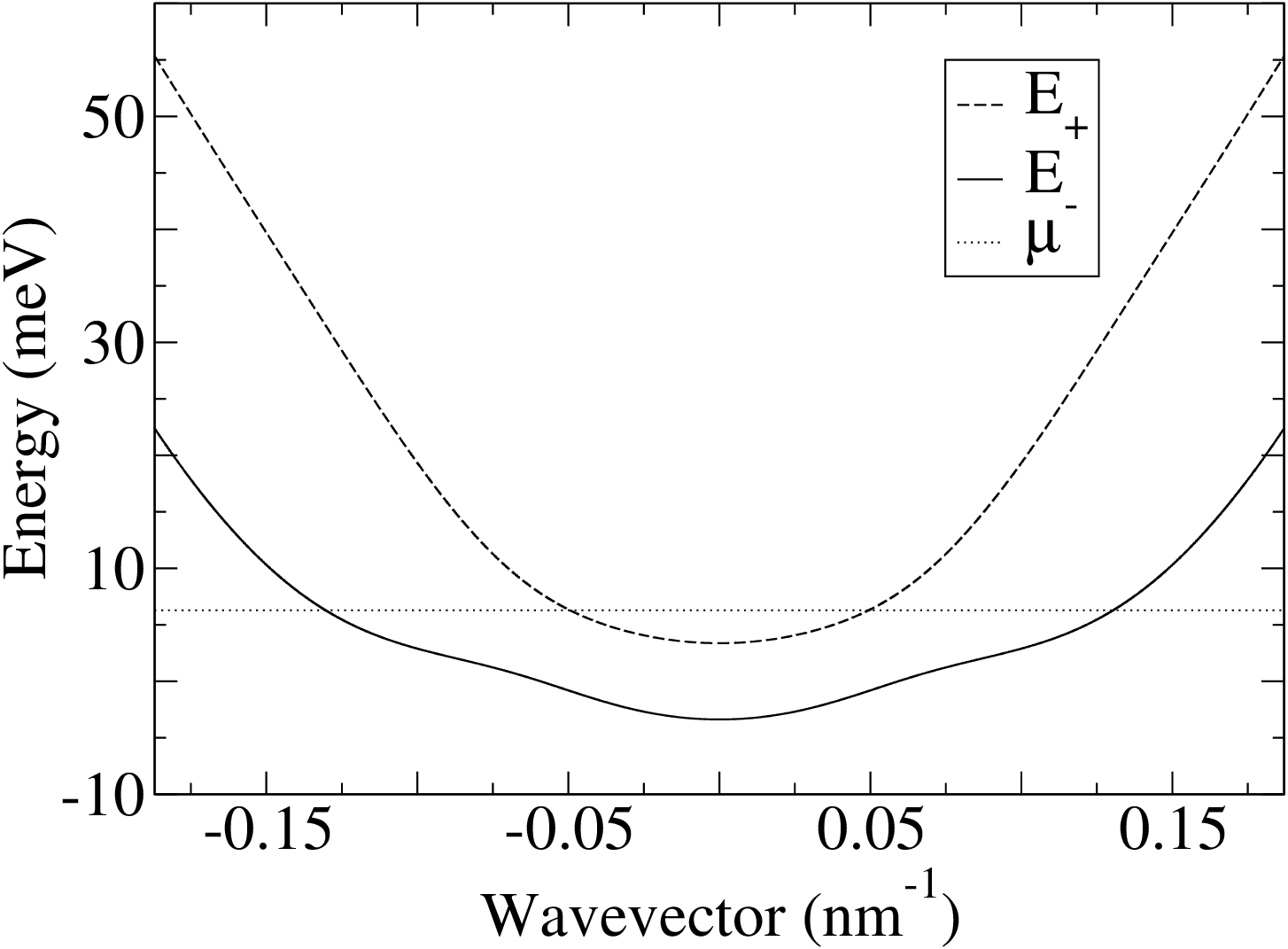}
\vspace{0.2cm}
\caption{Band dispersion relation for 2D electrons in a zincblende lattice. The parameters are n = 2.8$\times$10$^{11}$cm$^{-2}$, $\alpha_{zb}$ = 10000meV nm$^3$, h$_0$ = 3.38meV and m$^*$=0.034m$_0$.}
\label{fig: k3energies}
\end{center}
\end{figure}
\begin{figure}[h]
\begin{center}
\epsfxsize=2.5in
\epsffile{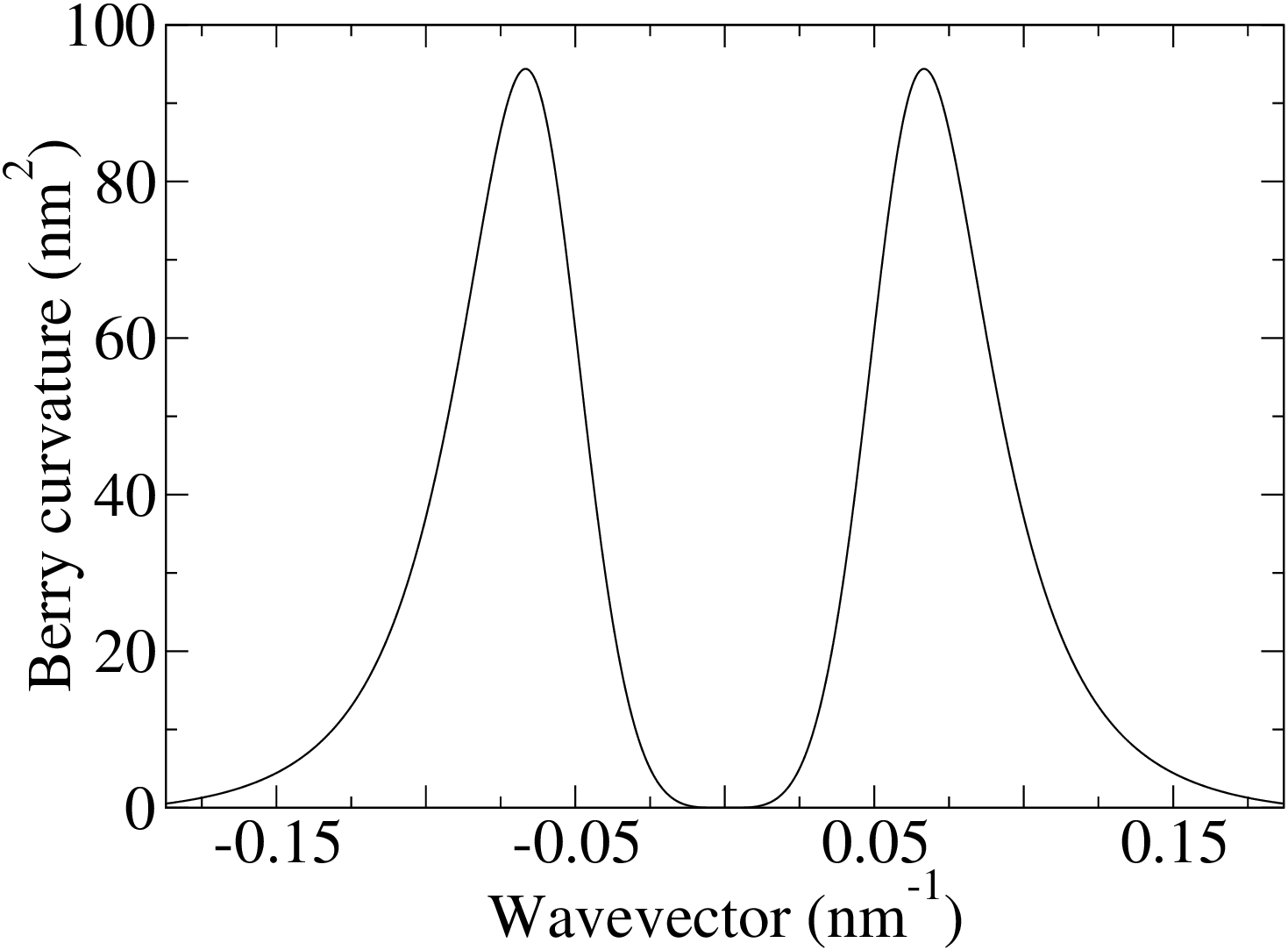}
\vspace{0.2cm}
\caption{Absolute value of the Berry curvature for the conduction band of zincblende structures.}
\label{fig: k3omega}
\end{center}
\end{figure}
\begin{figure}[h]
\begin{center}
\epsfxsize=2.5in
\epsffile{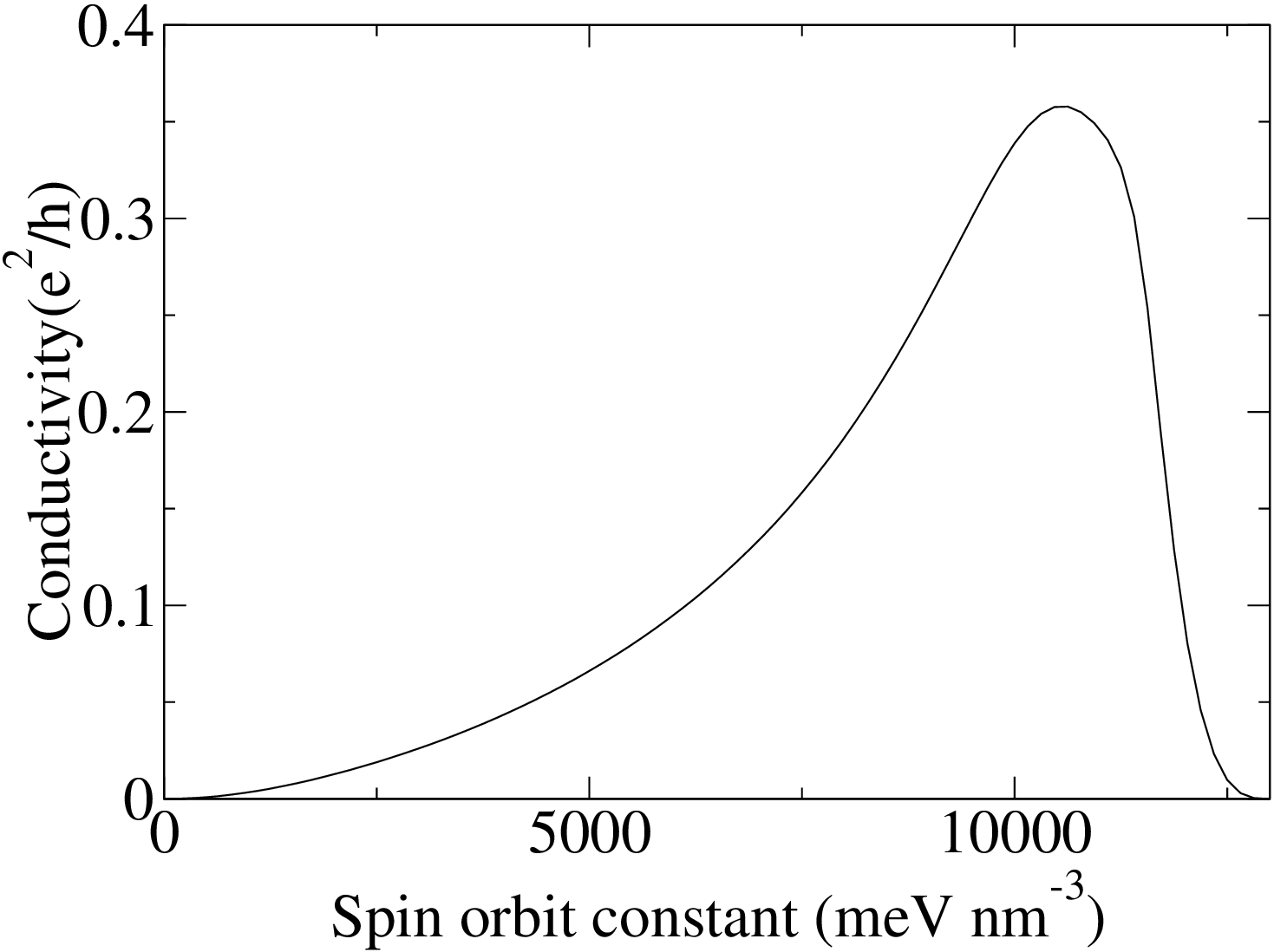}
\vspace{0.2cm}
\caption{Variation of the conductivity with the spin orbit constant for zincblende structures.}
\label{fig: k3betasigma}
\end{center}
\end{figure}
\begin{figure}[h]
\begin{center}
\epsfxsize=2.5in
\epsffile{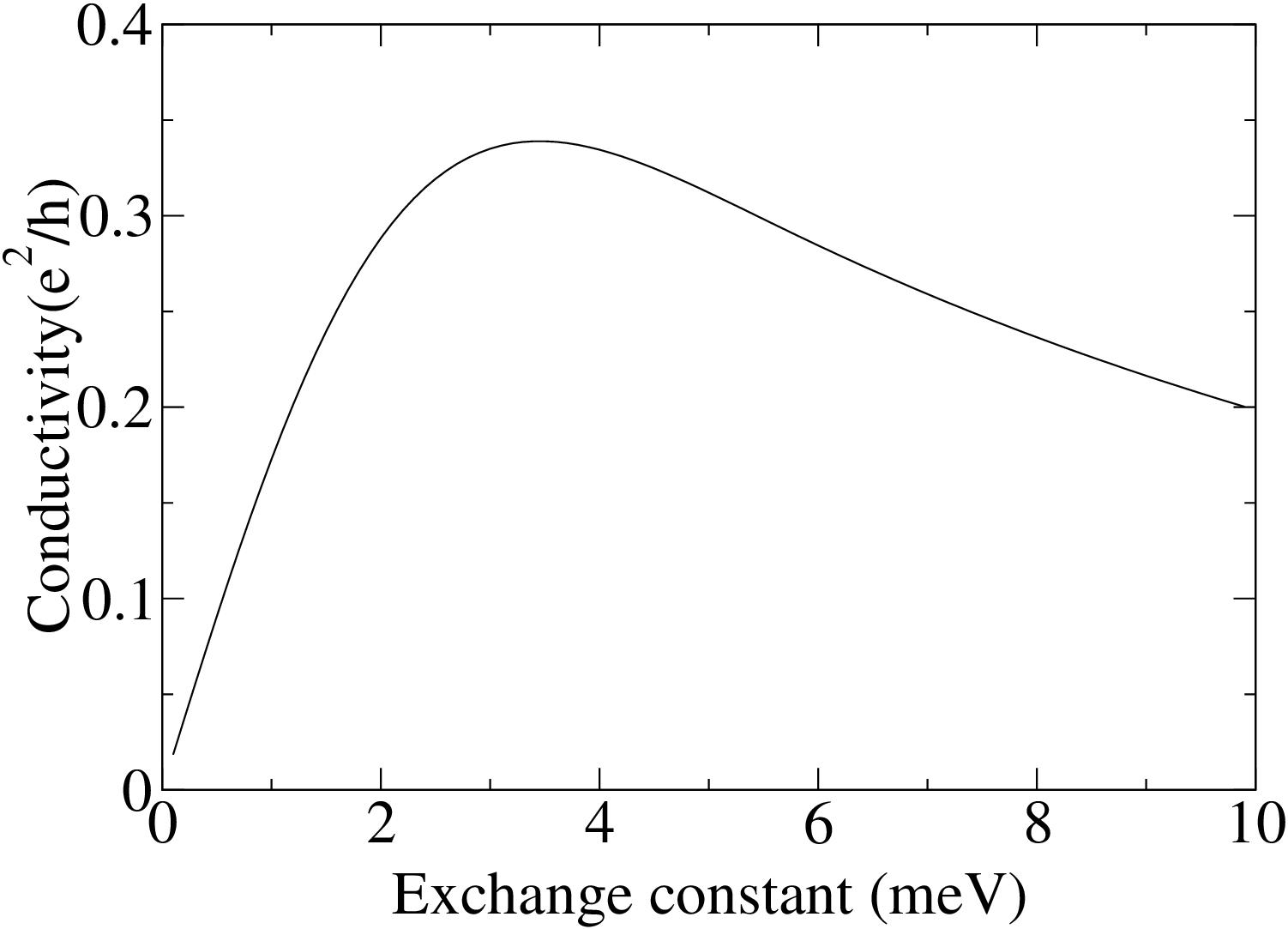}
\vspace{0.2cm}
\caption{Variation of the conductivity with the exchange field for zincblende structures. }
\label{fig: k3h0sigma}
\end{center}
\end{figure}
\begin{figure}[h]
\begin{center}
\epsfxsize=2.5in
\epsffile{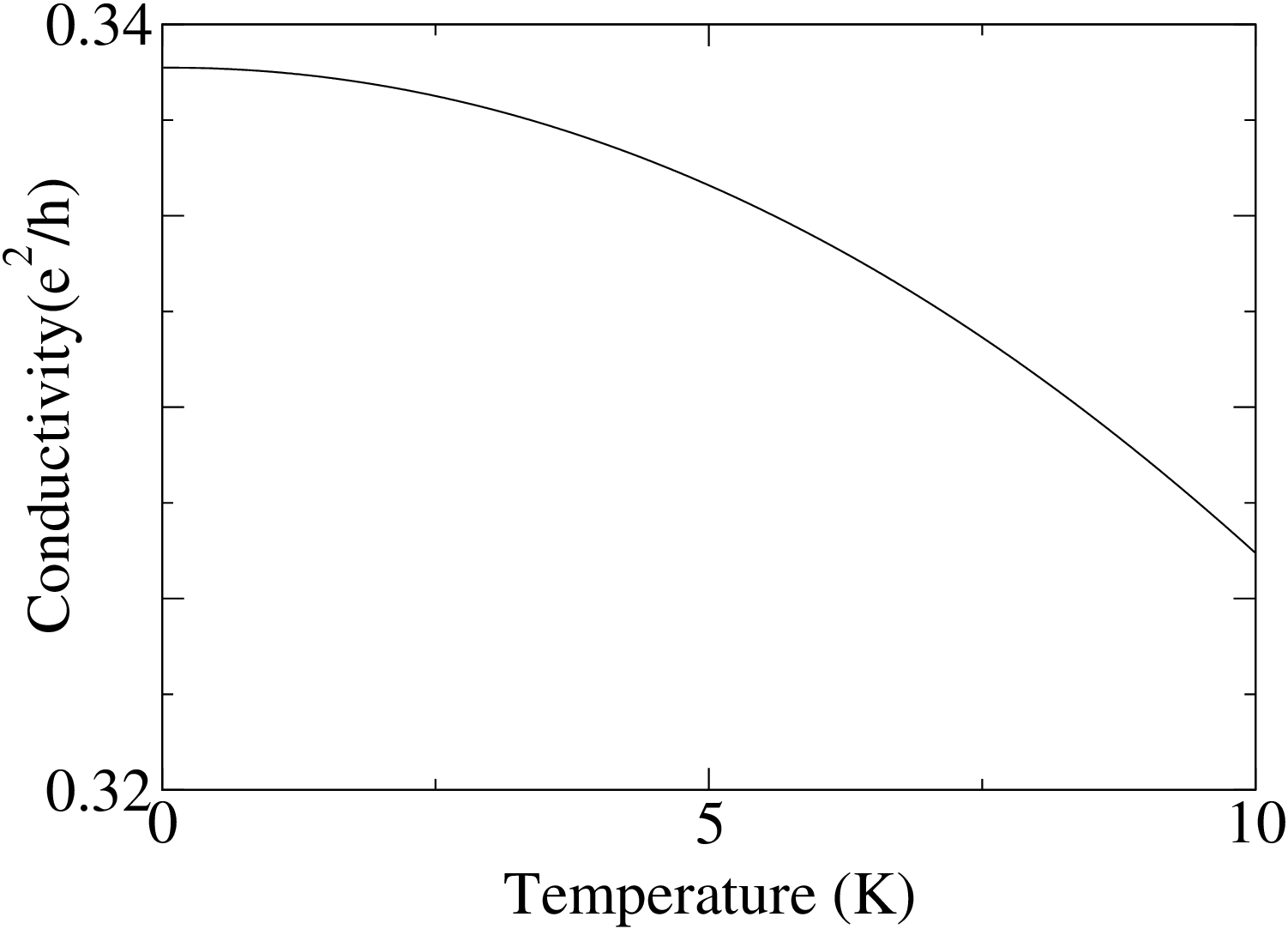}
\vspace{0.2cm}
\caption{Variation of the conductivity with temperature for zincblende structures.}
\label{fig: k3tsigma}
\end{center}
\end{figure}

\section{Other materials}
In general, the expression for $\alpha_{mat}$ is a$_{mat}\langle$E$\rangle$. Here $\langle$E$\rangle$ is the expectation value of the total electric field felt by the carriers and a$_{mat}$ a material specific parameter which is straightforwardly calculated using third order perturbation theory. It is customary to assume that the gradient of the confining potential has only a $z$ component, with the result that $\alpha_{mat}$ is given by a$_{mat}\langle$E$_z\rangle$. In the literature the size of the spin orbit coupling is parametrized either by direct measurements of $\alpha_{mat}$ (Table I), calculated values of a$_{mat}$ (Table II), or the magnitude of the energy splitting at k=0 or k=k$_F$. This disguises the fact that the character of E$_z$ is poorly understood and little literature is available on the topic. In a recent experiment, the electric field in the valence band of a GaAs/Ga$_{1-x}$Al$_x$As heterostructure was determined by Jusserand {\it et al}$^{22}$ to be 17 mVnm$^{-1}$. E$_z$ is assumed to scale with the band offset and can be increased by up to a factor of approximately 3.5 by applying a gate potential$^{23}$. In addition it was pointed out by Lassnig$^{24}$ that the conduction band spin splitting is due to the electric field in the valence band, the two fields differing through the contributions of the interfaces$^{24-25}$. 

We present in Table I the maximum observed/calculated values of $\alpha_{mat}$ in the bulk for semiconductors with a strong spin-orbit interaction (Table I) due to bulk inversion asymmetry (BIA). The small bulk GaAs spin-orbit coupling constant renders the effect of V$_{so}$ negligible in GaAs, but in the other materials the size of $\alpha_{mat}$ is several orders of magnitude larger.
\begin{table}
\begin{ruledtabular}
\begin{tabular}{ll}
GaAs$^{19}$ 	&  		0.69 \\
HgMnTe$^{26}$   & 		100 \\ 
InAs$^{27-28}$             &  		30-45 \\
HgTe gated QW$^{29}$   &  		40 \\
CdTe/HgTe/CdTe$^{30}$ &             40 \\ 
${\rm In_{0.75}Ga_{0.25}As/In_{0.75}Al_{0.25}As}$$^{31}$ &  		29.2 \\
${\rm In_{0.75}Ga_{0.25}As}$/InP$^{32-33}$ &  		4.71-15 \\
CdSe(holes)$^{34}$    & 	$<10$ (expt) \\
CdSe(holes)$^{34}$     &     6 (theory) \\
CdSe(electrons)$^{34}$    &    $<90$ (expt) \\
CdSe(electrons)$^{34}$    &     23 (theory) \\
\end{tabular}
\end{ruledtabular}
\end{table}
In Table II we list calculated values for the coefficient a$_{46}$ for 
different materials. By comparing with the corresponding values of $\alpha_{mat}$ one can obtain a rough estimate of the electric field in the valence band in the absence of a gate. In the case of InAs this field would lie in the range 25-40 meVnm$^{-1}$. 
\begin{table}
\begin{ruledtabular}
\begin{tabular}{ll}
GaAs$^{19}$             &  		0.055 \\
Hg$_{0.8}$Cd$_{0.2}$Te$^{19}$            & 		19.3 \\
InAs$^{25}$             &  		1.17 \\
InSb$^{25}$             &  		4-5.23 \\
ZnSe$^{25}$             &  		0.01 \\
\end{tabular}
\end{ruledtabular}
\end{table}

\section{Effect of magnetic field and disorder}
AHE was first observed in ferromagnetic materials in the absence of an 
external magnetic field (in this case one would only need to apply a field to magnetize the material, lowering it to zero afterwards). As ferromagnetic heterostructures are yet to be realized and as ferromagnetism has not been observed in II-VI semiconductors, it is more sensible to consider the case of a paramagnetic system, in which the exchange field can be maintained only by applying an external magnetic field. In order to determine the regime in which AHE can be observed in a weak magnetic field one needs to consider the fact that a magnetic field will cause the system to be quantized into Landau levels - where the semiclassical approximation is not valid, as well as give rise to the ordinary Hall effect. 

The first obstacle is circumvented by the presence of disorder in the sample, as the impurity scattering causes the Landau levels to broaden so that for a small enough magnetic field they overlap. The effect of disorder is parametrized by an impurity scattering time $\tau$, which in II-VI heterostructures is of the order of 0.1ps$^{36}$. To get round the second problem, the parameters must be matched so that the ordinary Hall conductivity does not overwhelm the anomalous one, making observation of the latter contribution clear. If the magnetic field and the scattering time are small enough to make the Landau levels overlap, $\omega_c\tau < 1$ must hold, where $\omega_c$ is the cyclotron frequency. The condition that $\omega_c\tau < 1$ ensures that the semiclassical approximation is valid, but does not guarantee that the ordinary Hall conductivity will not greatly exceed the anomalous one. For small $\omega_c\tau$ the ordinary Hall contribution, which, in the absence of quantum oscillations is given by the Drude formula: 
\begin{equation}
\sigma^{OHE}_{xy} = \frac{ne^2\tau}{m^*}\frac{\omega_c \tau}{1+\omega_c^2\tau^2}
\end{equation}
tends to zero. To ensure the AHE is the dominant effect we set:
\begin{equation}
\sigma^{OHE}_{xy} < \sigma^{AHE}_{xy}
\end{equation}
These two equations yield $(\frac{nh}{m^*}) \omega_c\tau^2 < 1$.

It is also imperative to ensure the AHE itself is not completely overshadowed
by disorder. To satisfy this requirement, the exchange splitting 
h$_0$ must exceed the energy fluctuation due to disorder, $\hbar \over\tau$. 
It follows that the condition for the observation of AHE is: 
\begin{equation}
\frac{2\pi n\hbar^2}{m^*}\omega_c\tau < \frac{\hbar}{\tau} < h_0
\end{equation}
For $\tau$=0.1ps, the fluctuation $\hbar \over\tau$ represents an energy of 6.5meV. 
 
As it is desired to work with a narrow well, so as to keep the subbands as far from each other as possible, we shall set the well width at 10nm, close to the smallest that can be manufactured. Furthermore, the laboratory temperature will be fixed at 0.1K. We use the exchange constants N$_0\alpha$ and N$_0\beta$ in table V of ref. 38 to determine the optimal Mn concentration and external magnetic field for the observation of the AHE in Cd$_{1-x}$Mn$_x$Se and Hg$_{1-x}$Mn$_x$Te quantum wells. 

For wurtzite (Cd$_{1-x}$Mn$_x$Se), with the value of $\alpha_{w}$ fixed we have chosen the carrier density n and exchange field h$_0$ in such a way as to have an observable conductivity in the valence band: n=2.9$\times$10$^{12}$ cm$^{-2}$ and  h$_0$=7meV. The Mn doping density will have to be 2.2$\%$. At 0.1K, in order for the Brillouin function to saturate the magnetic field must be approximately 1T. At this field, the ordinary Hall conductivity is less than 0.05 of the conductivity quantum, while the anomalous one is approximately 0.27.

In the case of zincblende (Hg$_{1-x}$Mn$_x$Te), the act of balancing the ordinary and anomalous conductivities is more difficult. The magnetic field cannot be as high as 1T, for that will produce a large ordinary contribution, but that is compensated by the fact that the exchange constant N$_0\beta$ is larger. 
In order to maintain the exchange splitting above the disorder broadening, i.e. h$_0$=7meV, it is sufficient to apply B=130mT and keep the Mn density unchanged at 2.2$\%$ (corresponding to 3.3$\times$10$^{26}$ m$^{-3}$, which is well within the experimentally achievable range$^{38}$). At a carrier density of 1$\times$10$^{11}$ cm$^{-2}$, the ordinary and anomalous conductivities will be equal to just over 0.14 of the conductivity quantum.  

This work was supported by the Department of Energy under Grant 
DE-FG03-02ER45958, and by the Welch Foundation in Texas.

\end{document}